\documentclass[%
reprint,
 superscriptaddress,
 amsmath,amssymb,
pra,
]{revtex4-2}

\usepackage{graphicx}
\usepackage{dcolumn}
\usepackage{bm}

\usepackage{comment}

\def\bra#1{\langle #1|}
\def\ket#1{|\mbox{$#1$}\rangle}
\def\bracket#1{\langle\mbox{$#1$}\rangle}

\def\bracketii#1#2#3{\langle\mbox{$#1$}|\mbox{$#2$}|\mbox{$#3$}\rangle}

\begin{document}

\preprint{APS/123-QED}

\title{Reconstruction of Wigner function of electron beams\\based on coherence measurements}

\author{Shuhei~Hatanaka}
\email{hatanaka@uhvem.osaka-u.ac.jp}
\affiliation{Research Center for Ultra-High Voltage Electron Microscopy, Osaka University, 7-1 Mihogaoka, Ibaraki, Osaka 567-0047, Japan}%
\affiliation{Graduate School of Engineering, Osaka University, 2-1 Yamadaoka, Suita, Osaka 565-0871, Japan}%
\author{Jun~Yamasaki}
\affiliation{Research Center for Ultra-High Voltage Electron Microscopy, Osaka University, 7-1 Mihogaoka, Ibaraki, Osaka 567-0047, Japan}%
\affiliation{Institute of Materials and Systems for Sustainability, Nagoya University, Furo-cho, Chikusa, Nagoya 464-8601, Japan}%
\date{\today}

\begin{abstract}
We developed a reconstruction method for the density matrix and Wigner function of electron beams through analysis of the Airy pattern intensity profile. The density matrix in a transmission electron microscope object plane was calculated using the coherence function and the electron wave amplitude and phase distributions. The Wigner function was then reconstructed using the matrix elements. Based on the Wigner function at the origin of the phase space, we derived a formula to calculate the axial brightness, and then determined the axial brightness of a Schottky field emission gun, which reflects the emitter performance more accurately and precisely than the conventional mean brightness measurements.
\end{abstract}

\maketitle


\section{INTRODUCTION}
Electron beams are important probes to measure material structures and electronic states on the nanometer scale, particularly in transmission electron microscopes (TEMs). Recently, quantum measurements of electron beams themselves and control of electron states have been realized in TEMs, e.g., controlling electron states via the interaction between free electrons and the optical near field\cite{Echternkamp2016}, measurement of antibunching of electrons\cite{Kuwahara2021}, and realization of quantum logic gates for free electrons\cite{Loffler2023}. The possibility of decoherence measurements of electrons entangled with bulk plasmon and surface plasmon is also proposed\cite{Mechel2021}.

Electrons generated from an electron emitter are in a mixed state\cite{Lubk2015semiclassical} that can be expressed using a density operator. The density operator contains all knowable information about the quantum system, including the wave nature e.g., the phase and coherence of the electron waves, which cannot be described via the particle model. For quantum measurements such as decoherence measurements, state measurements based on the density operators before and after the interaction are important. 

One alternative way to describe the information contained in the density operator is to use the Wigner function, which describes the state in phase space spanned by the position and momentum bases. When considered in a 2D plane, the Wigner function is given by\cite{Wigner1932}
\begin{equation}
W(\bm{r},\bm{q}) =  \frac{1}{(2 \pi)^2} \iint_{-\infty}^{\infty} d^2 \bm{\mu}\  e^{-i \bm{q} \bm{\mu}} \bracketii{\bm{r} + \frac{\bm{\mu}}{2}}{\hat{\rho}}{\bm{r} - \frac{\bm{\mu}}{2}}\label{wigner},
\end{equation}
where $\bm{r}$ and $\bm{\mu}$ are 2D real space vectors, and $\bm{q}$ is a 2D reciprocal space vector. $\hat{\rho}$ is the density operator. It is known that Wigner functions can have negative values as a manifestation of quantum nature, as measured for photons and laser cooled ions in Fock states\cite{Lvovsky2001,Leibfried1996}, and matter wave interference\cite{Kurtsiefer1997}. Because of this negative value characteristic, the Wigner function is a quasiprobability distribution.

The phase space representation of a beam in an optical system can also be derived in a classical manner using ray diagrams, but the Wigner function gives a complete description of the wave field, including the coherence, intensity distribution, and phase distribution in phase space. The Wigner function thus allows evaluation of the power going in each direction at each position in the optical system. Therefore, when compared with the density operator, the Wigner function is advantageous when evaluating the axial brightness in a TEM, which represents the electric current propagating along the optical axis. 

In quantum optics, the Wigner function has been reconstructed in various ways e.g., phase space tomography\cite{Raymer1994,Mcalister1995}, optical homodyne tomography\cite{Smithey1993}, and heterodyne measurement\cite{Lee1999}. It is, however, difficult to apply these methods to actual electron waves because of the lack of flexibility of the TEM optical system. To reconstruct the Wigner functions of electron waves, methods based on in-line or off-axis holography in TEMs have been investigated theoretically\cite{Lubk2015phase}. However, the limitations on the variable range of the illumination lens system and Fresnel scattering by the biprism cause serious problems when attempting to obtain the correct reconstructions in each method\cite{Lubk2015phase}.

Here, we propose a method to measure the density matrix based on the Airy pattern from an aperture. Using the matrix elements measured for the electron beams in a TEM, the Wigner function of the electron waves is reconstructed for the first time. In addition, using the reconstructed Wigner function, the axial brightness, which is an important performance indicator for emitters, is measured with greater accuracy and precision than in conventional measurements.

\section{THEORY AND METHOD}
\begin{figure}[htbp]
\centering\includegraphics[width=6.5cm]{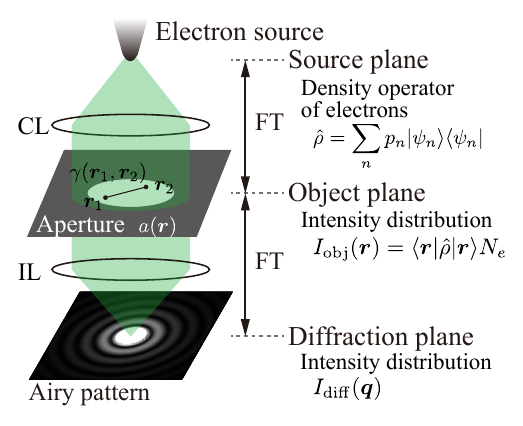}
\caption{Simplified drawing of the optical system to form the Airy pattern in a TEM. CL, condenser lens system; IL, imaging lens system; FT, Fourier transform relation; $\gamma(\bm{r}_1,\bm{r}_2)$, coherence function; $a(\bm{r})$, aperture function; $\hat{\rho}$, density operator.}
\label{opticalSys}
\end{figure}

Generally, the diagonal density matrix elements are obtained via intensity measurements, whereas phase measurements are performed to determine the off-diagonal elements. In our previous research, a method to determine the phase distribution in real space via Airy pattern intensity analysis was developed\cite{Yamasaki2018}. A simplified drawing of the optical diagram in a TEM to record the Airy pattern is shown in Fig.~\ref{opticalSys}, where the illumination system composed of multiple condenser lenses is denoted by CL, and the imaging system composed of multiple intermediate lenses, a projection lens, and lenses in the energy filter is denoted by IL. As described in \cite{Yamasaki2018}, the purpose to use the energy filter is not to eliminate electrons with energy fluctuations but to increase the camera length enough to measure the Airy patterns with a sufficiently fine sampling interval.

In a TEM, electrons generated by an emitter pass through the CL and then illuminate the object plane. The beam intensity at the object plane is $I_{\mathrm{obj}}(\bm{r})=\xi (\bm{r})^2$, where $\xi (\bm{r})$ is the amplitude of the wave field $\Psi(\bm{r})$. As Fig.~\ref{opticalSys} shows, when a circular aperture in the object plane is illuminated using a nearly parallel beam, an Airy pattern appears in the diffraction plane formed by the IL. Because all electron sources have a finite size, then based on the particle model, electrons emitted from different points on the source surface reach the object plane from slightly different directions.

Using the wave model, this situation can be expressed as an incoherent superposition of the electrons in different states. The statistical mixture of these different states, i.e., the mixed state is expressed using the density operator $\hat{\rho} = \sum_n p_n \ket{\psi_n}\bra{\psi_n}$, where $p_n$ is the probability of the state $\ket{\psi_n}$. To be precise, energy fluctuations between the electrons are also represented by different states. Because the energy spread of electrons emitted from a Schottky FEG is approximately 1~eV, the relative fluctuation in the 200-keV beam is in the order of $5\times10^{-6}$. 
The relative fluctuation is so small that the chromatic aberration and the temporal coherence do not affect the Airy pattern which is the small-angle scattering from the aperture\cite{Yamasaki2018,Hatanaka2021}. This means that the energy spread can be neglected for the electron wave inside the aperture. Under the monochromatic approximation, the electron wavelength $\lambda=2.51\,\mathrm{pm}$ at the acceleration voltage of 200~kV.

In quantum mechanics, a two-point correlation is expressed using $\bracketii{\bm{r}_1}{\hat{\rho}}{\bm{r}_2}$ with the normalization $\mathrm{Tr}[\hat{\rho}]=1$ for $\bm{r}$ inside the aperture. In optics, the two-point correlation of a wave field is expressed using the mutual coherence function $\Gamma(\bm{r}_1,\bm{r}_2)=\bracket{\Psi(\bm{r}_1)\Psi^{*}(\bm{r}_2)}$, where $\bracket{}$ is the ensemble average\cite{Raymer1994}. $\Gamma(\bm{r},\bm{r})$ is normalized via the integration within the aperture area $S_{\mathrm{a}}$ as $\iint_{S_{\mathrm{a}}}\ d^2\bm{r} \Gamma(\bm{r},\bm{r}) = \iint_{S_{\mathrm{a}}} d^2 \bm{r}\ \xi(\bm{r})^2 = N_e$, where $N_e$ is the number of electrons forming the Airy pattern.

When the CL is adjusted ideally to realize the Fourier transform relationship between the emitter plane and the object plane, the parallel illumination $\Psi(\bm{r})$ with a uniform amplitude $\xi_0$ and a uniform phase in the object plane is realized. By considering the difference of the normalization conditions in the mutual coherence function and in the density matrix, the off-diagonal elements of the density matrix are written as 
\begin{align}
\bracketii{\bm{r}_1}{\hat{\rho}}{\bm{r}_2} &= \Gamma(\bm{r}_1,\bm{r}_2)/N_e
= \gamma(\bm{r}_1,\bm{r}_2) \xi_0^2/N_e,\label{dens}
\end{align}
where $\gamma(\bm{r}_1, \bm{r}_2) := \Gamma(\bm{r}_1, \bm{r}_2)/\sqrt{\Gamma(\bm{r}_1, \bm{r}_1)\Gamma(\bm{r}_2, \bm{r}_2)}$ is called the coherence function, the absolute value of which represents the degree of coherence\cite{Born1999}.

In most practical cases, the illumination beam is not parallel, but is more or less converging/diverging on the object plane, which is expressed as a defocusing effect of the CL. The defocus aberration changes the beam diameter on the object plane, thus causing an increase/reduction of the average amplitude value from $\xi_0$. More generally, the amplitude in an actual illumination beam is described as $\xi(\bm{r})$, because other aberrations (and other practical reasons, e.g., slight beam misalignment from the optical axis) may induce an amplitude distribution. The actual illumination beam has also the phase distribution induced by the CL aberration, as can be inferred from the fact that a curved wave front is formed in a converging or diverging beam. These phase modulations are generally expressed by applying the unitary operator $\hat{U}_{\mathrm{CL}}$ to the electron states. Therefore, the state after the phase shift is expressed using $\ket{\psi'}=\hat{U}_{\mathrm{CL}} \ket{\psi}$. $\hat{U}_{\mathrm{CL}}$ has the eigenvalue $e^{-i k X_{\mathrm{CL}}(\bm{r})}$ for the position basis; $\hat{U}_{\mathrm{CL}} \ket{\bm{r}}=e^{-i k X_{\mathrm{CL}}(\bm{r})} \ket{\bm{r}}$, where $X_{\mathrm{CL}}(\bm{r})$ and $k=2\pi/\lambda$ are the axial geometric aberration of the CL and the wavenumber, respectively. The density operator after the influence of the lens aberration is expressed by $\hat{\rho}'=\sum_n p_n \hat{U}_{\mathrm{CL}}\ket{\psi_n}\bra{\psi_n}\hat{U}_{\mathrm{CL}}^{\dagger}$. Considering $\hat{U}_{\mathrm{CL}}^{\dagger} \ket{\bm{r}}=e^{ik X_{\mathrm{CL}}(\bm{r})} \ket{\bm{r}}$ and replacing $\xi_0$ in Eq.~(\ref{dens}) with $\xi(\bm{r})$, the off-diagonal elements of the density matrix are calculated using:
\begin{align}
\bracketii{\bm{r}_1}{\hat{\rho}'}{\bm{r}_2} =& \sum_n p_n \bracketii{\bm{r}_1}{\hat{U}_{\mathrm{CL}}}{\psi_n} \bracketii{\psi_n}{\hat{U}^{\dagger}_{\mathrm{CL}}}{\bm{r}_2}\notag\\
=& \ e^{-i k (X_{\mathrm{CL}}(\bm{r}_1)-X_{\mathrm{CL}}(\bm{r}_2))} \gamma(\bm{r}_1, \bm{r}_2) \xi(\bm{r}_1) \xi(\bm{r}_2)/N_e.\label{densityOp}
\end{align}

\begin{figure}[htbp]
\centering\includegraphics[width=7.5cm]{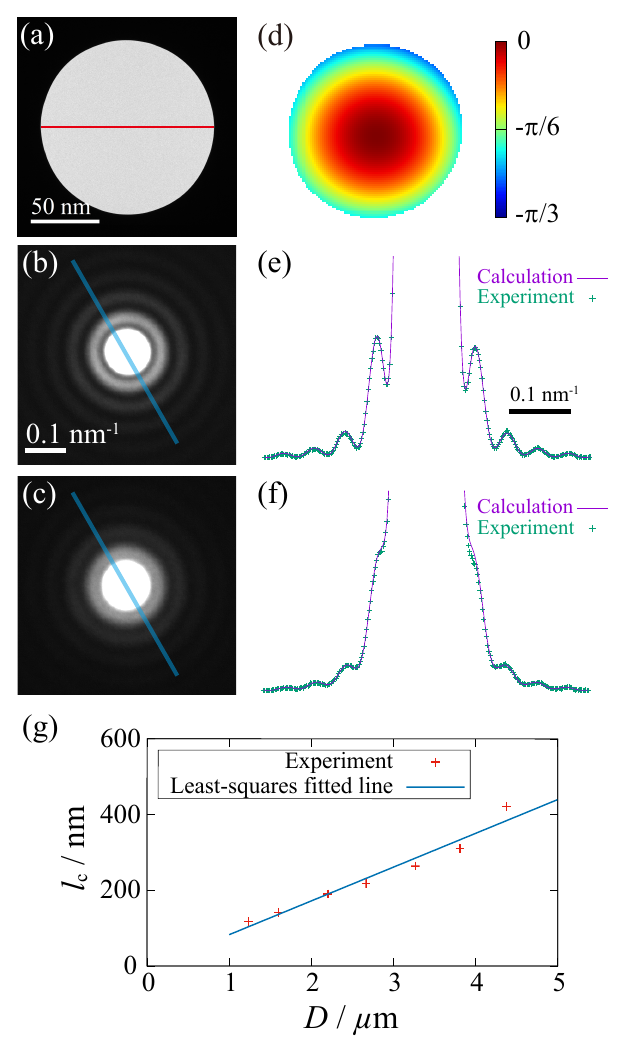}
\caption{Examples of the fitting analysis for the Airy patterns. (a) TEM image of the selector aperture. Airy patterns obtained using beams with (b) diameter $D = 4.4\,\mathrm{\mu m}$ and (c) $D = 1.2\,\mathrm{\mu m}$. (d) Phase distribution inside the aperture determined by the fitting calculation to the Airy pattern in (c). Comparison between the measured and calculated patterns along the solid lines in (b) and (c) are shown in (e) and (f), respectively. (g) Coherence length $l_{\mathrm{c}}$ as a function of beam diameter $D$. The diffraction data in (b) and (c) were selected as examples to illustrate the fitting analysis procedure from the data presented in \cite{Yamasaki2018}. The graph in (g) was reproduced from Fig.~7(a) in \cite{Yamasaki2018}.}
\label{Airyresult}
\end{figure}

Similar to the way in which the electron waves are modified by the CL aberration before reaching the object plane, a practical Airy pattern is influenced again by the axial geometric aberration of the IL, $X_{\mathrm{IL}}(\bm{r})$\cite{Yamasaki2018,Hatanaka2021}. Therefore, the electrons that form the Airy pattern are modulated by the sum of these aberrations: $X(\bm{r})=X_{\mathrm{CL}}(\bm{r})+X_{\mathrm{IL}}(\bm{r})$. When the object plane is illuminated via wave packets coming from different source positions, the Airy pattern is blurred with an angular distribution. The van Cittert–Zernike theorem indicates that the angular distribution has a Fourier transform relationship with $\gamma(\bm{r}_1,\bm{r}_2)$\cite{Born1999}. Therefore, the practical Airy pattern intensity after excluding electric and mechanical instabilities in the instrument is described as\cite{Yamasaki2018}:
\begin{align}
I_{\mathrm{diff}}(\bm{q}) = \left| \mathcal{F} \left[ a(\bm{r}) \xi(\bm{r}) e^{-i k X(\bm{r}) } \right] \right|^2 \otimes \mathcal{F}[\gamma(\bm{r}_1, \bm{r}_2)],\label{AiryFitting}
\end{align}
where $\mathcal{F}$ and $\otimes$ represent the Fourier transform and convolution operations, respectively. $a(\bm{r})$ is an aperture function that takes a value of 1 inside and 0 outside. Since $\xi(\bm{r})$, $X(\bm{r})$, and $\gamma(\bm{r}_1,\bm{r}_2)$ can be approximately expressed as parameterized functions\cite{Yamasaki2018}, these parameters are determined by the fitting calculation to the measured Airy patterns using Eq.~(\ref{AiryFitting}), combined with $a(\bm{r})$ determined from a TEM image of the aperture. The fitting procedure is described in detail in \cite{Yamasaki2018} and briefly in this paper using examples in Fig.~\ref{Airyresult} and Table~\ref{AberrationCoefficient}, later. 

Using the determined $\xi(\bm{r})$ and $\gamma (\bm{r}_1, \bm{r}_2)$, and by replacing $X_{\mathrm{CL}}(\bm{r})$ in Eq.~(\ref{densityOp}) with the determined $X(\bm{r})$, the off-diagonal elements are then calculated. Note that because of the difference between $X_{\mathrm{CL}}(\bm{r})$ and $X(\bm{r})$, the result is not for the actual wave in the object plane, but is for a virtual wave that includes the additional phase shift of $k X_{\mathrm{IL}}(\bm{r})$. The diagonal elements also can be calculated using Eq.~(\ref{densityOp}) as $\bracketii{\bm{r}}{\hat{\rho}'}{\bm{r}}=\xi(\bm{r})^2/N_e$, from the determined $\xi(\bm{r})$ by the fitting calculation. By substituting Eq.~(\ref{densityOp}) as described using $X(\bm{r})$ into Eq.~(\ref{wigner}), the Wigner function reconstructed within the aperture $W_{\mathrm{a}} (\bm{r},\bm{q})$ is described using
\begin{widetext}
\begin{align}
W_{\mathrm{a}} (\bm{r},\bm{q}) &=  \frac{1}{(2 \pi)^2 N_e} \iint_{-\infty}^{\infty} d^2 \bm{\mu}\ e^{-i \bm{q}\bm{\mu}}\ e^{-i k (X(\bm{r}+\frac{\bm{\mu}}{2})-X(\bm{r}-\frac{\bm{\mu}}{2}))}\ \gamma(\bm{r}+\frac{\bm{\mu}}{2}, \bm{r}-\frac{\bm{\mu}}{2}) a(\bm{r}+\frac{\bm{\mu}}{2}) \xi(\bm{r}+\frac{\bm{\mu}}{2}) a(\bm{r}-\frac{\bm{\mu}}{2}) \xi(\bm{r}-\frac{\bm{\mu}}{2})\label{wigner2}.
\end{align}
\end{widetext}

The analysis in this paper was conducted using experimental data that were previously reported in \cite{Yamasaki2018}. Brief descriptions of the measurement procedure and results are given as follows. A 200-kV TEM equipped with a Schottky field emission gun (FEG) (JEM-ARM200F, JEOL) was used. A selector aperture (SA) installed in the TEM with an effective diameter of 127~nm at the object plane (Fig.~\ref{Airyresult}(a)) was used to form the Airy patterns. $a(\bm{r})$ is created by binarizing Fig.~\ref{Airyresult}(a). Airy patterns are obtained with various beam diameters at the object plane to vary the spatial coherence inside the aperture from partially coherent to almost fully coherent. Figures~\ref{Airyresult}(b) and (c) show examples of the measured Airy patterns.

Under an assumption that the electron source intensity has a 2D Gaussian distribution, the 2D intensity profiles in Figs.~\ref{Airyresult}(b) and (c) are well reproduced by the fitting calculations. Table~\ref{AberrationCoefficient} shows the main parameters determined by the fitting calculations. An example of the phase distribution reconstructed using the aberration coefficients, $C_1$, $A_1$, $A_2$ and $B_2$ for the beam diameter $D=1.2\,\mathrm{\mu m}$ is shown in Fig.\ref{Airyresult}(d). As representative of these results, comparisons along the lines in Figs.~\ref{Airyresult}(b) and (c) are shown in Figs.~\ref{Airyresult}(e) and (f), respectively. As shown in Fig.~\ref{Airyresult}(g), the coherence length $l_{\mathrm{c}}$ as the standard deviation of $\gamma(\bm{r}_1,\bm{r}_2)$ shows a linear relationship with $D$ at the object plane, which was predicted theoretically\cite{Pozzi1987}. The proportional constant depends on the electron emitters, illumination lens setting, and the CL aperture size\cite{Yamasaki2018,Hatanaka2021}.

\begin{table}[hbtp]
  \caption{Examples of the main parameters determined by the fitting calculation to the Airy patterns obtained by beams with $D=4.4\,\mathrm{\mu m}$ shown in Fig.~\ref{Airyresult}(b) and $D=1.2\,\mathrm{\mu m}$ shown in Fig.~\ref{Airyresult}(c). $C_1$, $A_1$, $A_2$ and $B_2$ are defocus, two-fold astigmatism, three-fold astigmatism, and axial coma aberration coefficients, respectively. Because these are the aberration coefficients to form the Airy patterns in reciprocal space, they have dimensions of reciprocal length. $l_{\mathrm{c}}$ is coherence length, which is the standard deviation of $\gamma(\bm{r}_1, \bm{r}_2)$. $\Delta I_x$ and $\Delta I_y$ represent the beam intensity gradients in the $r_x$ and $r_y$ directions, respectively, expressed by ratios to the average intensity inside the aperture.}
  \label{AberrationCoefficient}
  \centering
  \begin{tabular}{l|ll}
    \hline
    parameter & $D=4.4\,\mathrm{\mu m}$ & $D=1.2\,\mathrm{\mu m}$ \\
    \hline
    $C_1 / \mathrm{m^{-1}}$ & -87.3  &  -138\\
    $|A_1| / \mathrm{m^{-1}}$  & 8.44 & 8.60\\
    argument of $A_1$ / deg. & -102 & 169\\
    $|A_2| / \mathrm{m^{-1}}$  & 0.0925 & 0.0641\\
    argument of $A_2$ / deg. & -134 & -160\\
    $|B_2| /\mathrm{m^{-1}}$ & 0.0161 & 0.0600\\
    argument of $B_2$ / deg. & 43.8 & 77.3\\
    $l_{\mathrm{c}} / \mathrm{nm}$ & 422 & 117\\
    $\Delta I_x /\mathrm{nm^{-1}}$ & $-1.20 \times 10^{-4}$ & $-1.34 \times 10^{-3}$ \\
    $\Delta I_y /\mathrm{nm^{-1}}$ & $-1.08 \times 10^{-4}$ &  $2.47 \times 10^{-4}$\\
    \hline
  \end{tabular}
\end{table}

\section{RESULTS AND DISCUSSION}
\begin{figure}[htbp]
\centering\includegraphics[width=7.5cm]{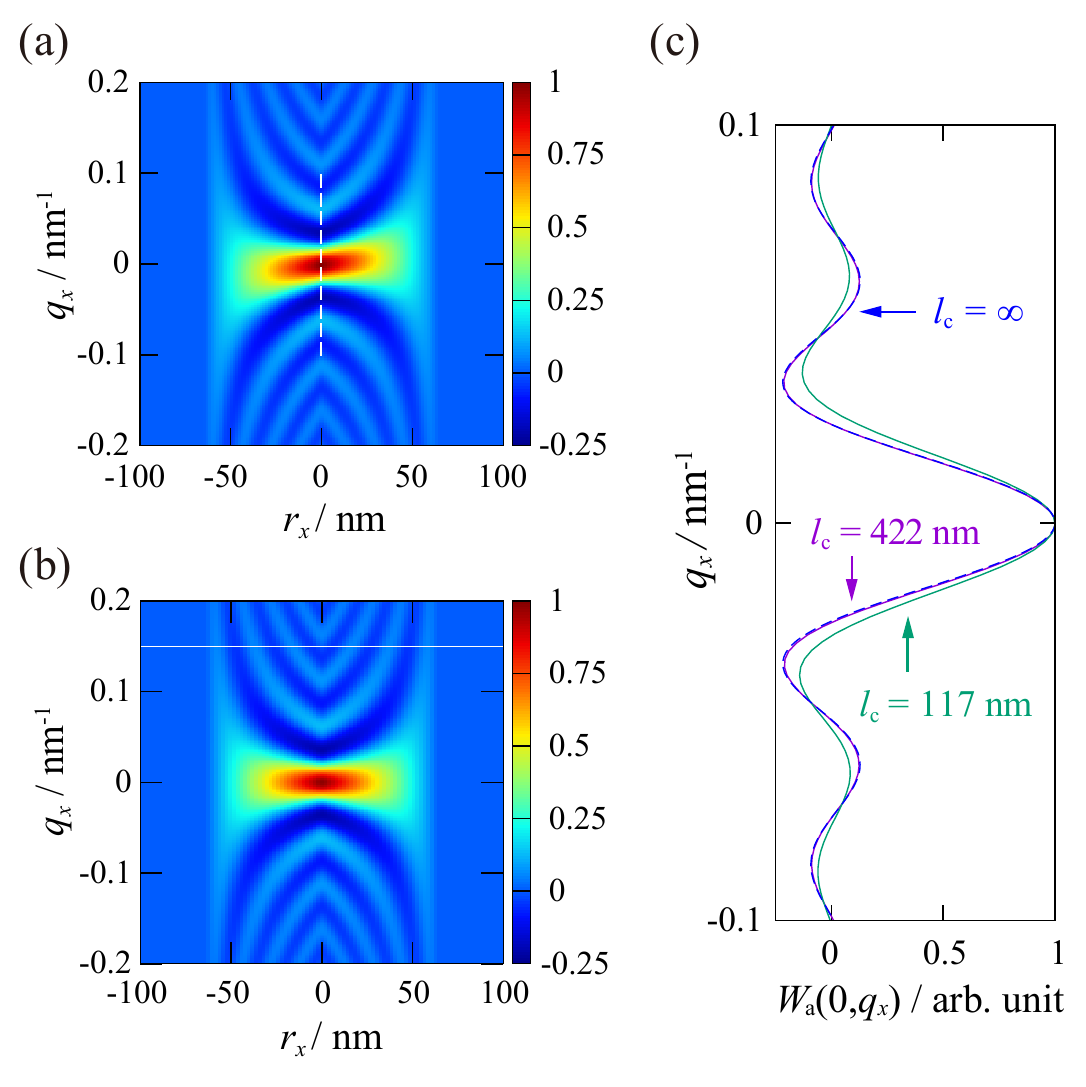}
\caption{Reconstruction of the Wigner function. (a) $W_{\mathrm{a}}(r_x, q_x)$ for the beam with $l_{\mathrm{c}} =422\,\mathrm{nm}$ calculated from the density matrix along the red line shown in Fig.~\ref{Airyresult}(a). The density matrix is calculated from the fitting calculation results of Fig.~\ref{Airyresult}(b). The values of $W_{\mathrm{a}}(r_x, q_x)$ when the main peak value is normalized to 1 are shown by color. (b) $W_{\mathrm{a}}(r_x, q_x)$ calculated for an ideal beam with infinite coherence length and without lens aberrations and amplitude distribution. (c) Comparison along the broken line in (a) between $W_{\mathrm{a}}(0, q_x)$ for the beams with $l_{\mathrm{c}}=422\,\mathrm{nm}$ and $117\,\mathrm{nm}$, and the ideal beam with $l_{\mathrm{c}}=\infty$.}
\label{WignerSchottky}
\end{figure}

Using the parameters determined from the Airy patterns, the four-dimensional Wigner functions $W_{\mathrm{a}}(\bm{r},\bm{q})$ for the beams with various $l_{\mathrm{c}}$ can be reconstructed from Eq.~(\ref{wigner2}). An example of the visualized 2D phase space $W_{\mathrm{a}}(r_x, q_x)$ is shown in Fig.~\ref{WignerSchottky}(a), which was reconstructed for the beam with $l_{\mathrm{c}} = 422\ \mathrm{nm}$ using the density matrix $\bracketii{r_{x1}}{\hat{\rho}'}{r_{x2}}$ relating to the 1D position basis $r_x$ along the line in Fig.~\ref{Airyresult}(a). Projections of the values of $W_{\mathrm{a}}(r_x,q_x)$ onto the $r_x$ axis corresponds to the intensity profile along the line in Fig.~\ref{Airyresult}(a). The oscillation in $W_{\mathrm{a}}(r_x, q_x)$ reflects the fringe in the Airy pattern. Therefore, the negative values in $W_{\mathrm{a}}(r_x, q_x)$ are caused by the diffraction phenomenon of the electron waves passing through the aperture. For comparison, $W_{\mathrm{a}}(r_{x}, q_{x})$ of the virtual ideal beam was calculated using uniform amplitude $\xi(\bm{r})=1$, phase $X(\bm{r})=0$, and $\gamma(\bm{r}_1, \bm{r}_2 )=1$. As shown in Fig.~\ref{WignerSchottky}(b), the overall patterns are similar to the experiment in Fig.~\ref{WignerSchottky}(a). The difference is a slightly tilted crest of high values (red region) around the origin in the experiment. By analogy from the phase space representation based on ray diagrams and the phase space tomography\cite{Raymer1994,Mcalister1995}, such a tilted crest is induced by the curved wavefront in a diverging beam. In the present case, the tilted crest in Fig.~\ref{WignerSchottky}(a) reflects the sum of the defocus values of the CL and the IL because $W_{\mathrm{a}}(r_x, q_x)$ includes also the influence of the IL aberration. More generally, the overall deformation e.g., slight asymmetric feature in Fig.~\ref{WignerSchottky}(a) is induced by the remaining total aberrations of the CL and the IL.

Figure~\ref{WignerSchottky}(c) shows comparison of $W_{\mathrm{a}}(0, q_{x})$, that is, the profiles along the broken line in Fig.~\ref{WignerSchottky}(a) between the ideal beam and the experimental beams with $l_{\mathrm{c}} = 422\,\mathrm{nm}$ in Fig.~\ref{Airyresult}(b) and $117\,\mathrm{nm}$ in Fig.~\ref{Airyresult}(c). When the profiles are normalized by the peak values, there is no significant differences between $l_{\mathrm{c}}=422\ \mathrm{nm}$ and $l_{\mathrm{c}}=\infty$. This is because the electrons inside aperture with the diameter of 127~nm are almost fully coherent when illuminated by the beam with $l_{\mathrm{c}}=422\ \mathrm{nm}$. On the other hand, the  oscillation amplitude is slightly reduced in the profile for the beam with $l_{\mathrm{c}}=117\ \mathrm{nm}$. Considering the oscillation reflects the wave diffraction phenomenon as mentioned before, the amplitude reduction is attributed to reduction of the wave nature, that is, partial coherence of the beam inside the aperture. This is consistent with the fact that $l_{\mathrm{c}}=117\ \mathrm{nm}$ is slightly smaller than the aperture diameter of 127~nm.

As mentioned in the introduction, it is expected that important finding or knowledge is obtained based on the relationship between the Wigner function and the brightness, which is known to be an important indicator of the emitter performance. Specifically, the axial brightness $B_0$ is of primary importance in optical systems using lenses because the value of $B_0$ is conserved along the optical axis\cite{hawkes1989}. $B_0$ is defined by the electric current passing through an infinitely small area within an infinitely small solid angle along the optical axis, i.e., it is defined by the current density at the origin of the phase space \cite{Lubk2015semiclassical}.

Because small apertures of finite sizes are used in practical measurements\cite{hawkes1989}, the current is measured in an area called emittance that is not infinitely small. Division of the current value by the emittance gives $B_0$ correctly if the current density within the emittance is uniform, but the value is smaller than $B_0$ in reality because it has a nonuniform distribution with a peak at the origin of the phase space. This is called the mean brightness $\bar{B}$\cite{hawkes1989}. Moreover, because the size of the emittance depends on the lens setting and the aperture size used to perform the measurement\cite{hawkes1989}, measured $\bar{B}$ values tend to fluctuate depending on the TEM used and the researchers who conducted the measurements.

\begin{figure}[htbp]
\centering\includegraphics[width=6.5cm]{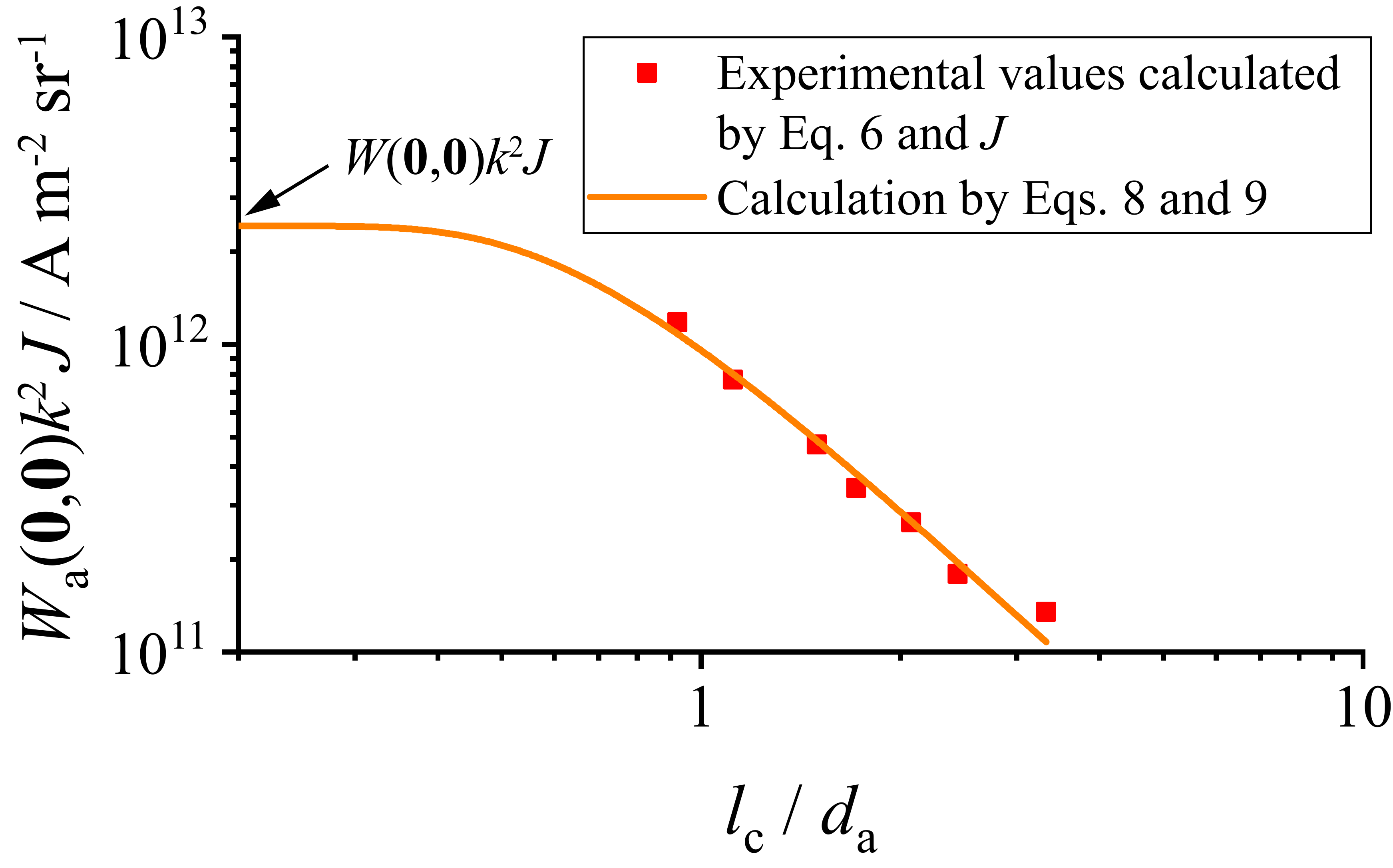}
\caption{Electric current at the origin of the phase space $W_{\mathrm{a}}(\bm{0},\bm{0})k^2 J$ as a function of $l_{\mathrm{c}}/d_{\mathrm{a}}$. $l_{\mathrm{c}}$ and $d_{\mathrm{a}}$ are the coherence length and the aperture diameter, respectively. The plotted dots are calculated using experimentally measured $W_{\mathrm{a}}(\bm{0}, \bm{0})$ and $J$ listed in Table~\ref{WaTable}. The curved line is drawn for Eq.~(\ref{wignerorigin}) using the $B_0$ value determined experimentally based on Eq.~(\ref{B0}). The value at $l_{\mathrm{c}}/d_{\mathrm{a}} \ll 1$ corresponds to the axial brightness.}
\label{CurrentOri}
\end{figure}

\begin{table}[hbtp]
  \caption{$W_{\mathrm{a}}(\bm{0},\bm{0})$ calculated by Eq.~(\ref{Wa00woapprox}) using the actual $\xi(\bm{r})$, $X(\bm{r})$, and $\gamma(\bm{r}_1, \bm{r}_2)$ and using uniform $X(\bm{r})=0$. The right column displays $J$ for each Airy pattern measured by a Faraday cup.}
  \label{WaTable}
  \centering
  \begin{tabular}{l|ll|l}
    \hline
    & \multicolumn{2}{c|}{$W_{\mathrm{a}}(\bm{0},\bm{0}) k^2 J/ \mathrm{A m^{-2} sr^{-1}}$} & \\
    $l_{\mathrm{c}} /\mathrm{nm}$ & actual $X(\bm{r})$ & $X(\bm{r}) = 0$ & $J/\mathrm{pA}$ \\
    \hline
    117 & $1.18 \times 10^{12}$ & $1.18 \times 10^{12}$ & 2.59 \\
    142 & $7.69 \times 10^{11}$ & $7.70 \times 10^{11}$ & 1.52 \\
    190 & $4.72 \times 10^{11}$ & $4.73 \times 10^{11}$ & 0.85 \\
    218 & $3.41 \times 10^{11}$ & $3.42 \times 10^{11}$ & 0.60 \\
    264 & $2.64 \times 10^{11}$ & $2.65 \times 10^{11}$ & 0.45\\
    310 & $1.79 \times 10^{11}$ & $1.79 \times 10^{11}$ & 0.30 \\
    422 & $1.35 \times 10^{11}$ & $1.35 \times 10^{11}$ & 0.22\\
    \hline
  \end{tabular}
\end{table}

Because the Wigner function gives the quasiprobability, $B_0$ as defined using the current density at the origin of the phase space is given by $W(\bm{0},\bm{0})k^2 J$\cite{Lubk2015semiclassical}, where $k^2$ is multiplied for the unit conversion from $\mathrm{nm}^{-2}$ to $\mathrm{sr}^{-1}$. $J$ is the total electric current in the phase space and, in this work, is equal to the electric current to form the Airy pattern. However, because of the aperture function in Eq.~(\ref{wigner2}), $W_{\mathrm{a}}(\bm{0},\bm{0})$ is different from $W(\bm{0},\bm{0})$. Therefore, an appropriate conversion from the measured $W_{\mathrm{a}}(\bm{0},\bm{0})$ to $W(\bm{0},\bm{0})$ is required to estimate $B_0$. 

From Eq.~(\ref{wigner2}),
\begin{align}
W_{\mathrm{a}}(\bm{0},\bm{0}) = &\frac{1}{(2 \pi)^2 N_e} \iint d^2 \bm{\mu}\ e^{-i k (X(\frac{\bm{\mu}}{2})-X(-\frac{\bm{\mu}}{2}))}\notag\\
& \gamma(\frac{\bm{\mu}}{2}, -\frac{\bm{\mu}}{2}) a(\frac{\bm{\mu}}{2}) \xi(\frac{\bm{\mu}}{2}) a(-\frac{\bm{\mu}}{2}) \xi(-\frac{\bm{\mu}}{2}).\label{Wa00woapprox}
\end{align}
If we consider an ideal situation that the amplitude and phase in the aperture are uniform, Eq.~\ref{dens}, Eq.~(\ref{Wa00woapprox}) becomes to
\begin{align}
W_{\mathrm{a}} (\bm{0},\bm{0}) = &\frac{\xi_0^{2}}{(2 \pi)^2 N_e}\iint d^2 \bm{\mu}\ a(\frac{\bm{\mu}}{2}) \gamma(\frac{\bm{\mu}}{2}, -\frac{\bm{\mu}}{2}),\label{wigner3}
\end{align}
for a circular aperture. According to the van Cittert-Zernike theorem\cite{Born1999}, $\gamma(\frac{\bm{\mu}}{2}, -\frac{\bm{\mu}}{2})$ is given by the Fourier transform of the electron source intensity distribution, which is often approximated to be a 2D Gaussian function\cite{Yamasaki2018}. Under this approximation, $\gamma(\frac{\bm{\mu}}{2}, -\frac{\bm{\mu}}{2})$ becomes to the 2D Gaussian function, $e^{-|\bm{\mu}|^2/2l_{\mathrm{c}}^2}$, which are used for the fitting analysis of the Airy patterns. Considering the Airy patterns are excellently reproduced in Fig.~\ref{Airyresult} and in \cite{Yamasaki2018}, the Gaussian approximation for $\gamma(\frac{\bm{\mu}}{2}, -\frac{\bm{\mu}}{2})$ should be reasonable for Schottky FEGs. Thus, Eq.~(\ref{wigner3}) is given by integration of $e^{-|\bm{\mu}|^2/2l_{\mathrm{c}}^2}$ within the range determined by the aperture diameter $d_{\mathrm{a}}$. Using the relationship $N_e = \xi_0^{2} S_{\mathrm{a}}$,
\begin{align}
W_{\mathrm{a}} (\bm{0},\bm{0}) k^2 J &\simeq \frac{k^2 J \xi_0^{2}}{\pi^2 N_e} 2\pi \int_{0}^{\frac{d_{\mathrm{a}}}{2}} d \left( \frac{\mu}{2}\right) \frac{\mu}{2} e^{-\frac{2(\mu/2)^2}{l_{\mathrm{c}}^2}}\notag\\
&=  B_0 \left( 1- e^{-d_{\mathrm{a}}^2/2l_{\mathrm{c}}^2} \right)\label{wignerorigin}\\
B_0 = \frac{k^2 J \xi_0^{2} l_{\mathrm{c}}^2}{2 \pi N_e}  &= 2 \pi \frac{J}{S_{\mathrm{a}}} \frac{l_{\mathrm{c}}^2}{\lambda^2} = 2 \pi j_0 \frac{l_{\mathrm{c}}^2}{\lambda^2},\label{B0}
\end{align}
where $\mu$ and $j_0$ are the norm of $\bm{\mu}$ and the axial current density, respectively. The prefactor in Eq.~(\ref{wignerorigin}) must be $B_0$, because $W_{\mathrm{a}}(\bm{0},\bm{0})k^2 J$ becomes $W(\bm{0},\bm{0})k^2 J$ if no aperture or an infinitely large aperture is used. In other words, $W_{\mathrm{a}}(\bm{0},\bm{0})k^2 J$ converges to $B_0$ when the ratio $l_{\mathrm{c}}⁄d_{\mathrm{a}}$ approaches zero, as depicted in Fig.~\ref{CurrentOri}. $B_0$ as given by Eq.~(\ref{B0}) is a function of $j_0$ and $l_{\mathrm{c}}$, which both vary with the CL setting, mainly D in the object plane changed by the defocus component of $X_{\mathrm{CL}}(\bm{r})$. As shown in Fig.~\ref{Airyresult}(g), $l_{\mathrm{c}}$ is a linear function of $D$. Considering that $j_0$ is inversely proportional to the square of $D$, the value of $j_0 l_{\mathrm{c}}^2$ and therefore $B_0$ given by Eq.~(\ref{B0}) is constant, regardless of the lens conditions, as expected for the axial brightness characteristics.

Equation~(\ref{wignerorigin}) was derived for a virtual ideal condition with uniform amplitude and phase. The influence of the actual non-uniform $\xi(\bm{r})$ and $X(\bm{r})$ are discussed as follows. It is generally known that the beam intensity at the object plane is rather uniform under a nearly parallel illumination in a Schottky FEG TEM, as seen in Fig.~\ref{Airyresult}(a). In fact, based on the fitting results\cite{Yamasaki2018}, the intensity distribution $\xi(\bm{r})^2$ inside the aperture is estimated to have deviations of only a few percent at most for all the beams used in the present study, as shown in  Table~\ref{AberrationCoefficient}. On the other hand, as shown in Fig.~\ref{Airyresult}(d), the phase shows not negligible amounts of deviation from a uniform distribution even if the Airy pattern is carefully focused for the recording. To examine the influence of the phase distribution, $W_{\mathrm{a}}(\bm{0}, \bm{0})$ calculated using actual $\xi (\bm{r})$, $X(\bm{r})$ and $\gamma(\bm{r}_1,\bm{r}_2)$ determined by the fitting analysis and $W_{\mathrm{a}}(\bm{0}, \bm{0})$ calculated using $X(\bm{r}) = 0$ are compared in Table~\ref{WaTable}. Considering the differences less than 1\% for all the beams, the influences of the actual phase distributions included in the Airy patterns used can be neglected. Thus, Eqs.~(\ref{wigner3})--(\ref{B0}) should be valid not only for the ideal parallel illumination but even for the data set used in the present study.


The $j_0$ value for each $D$ needed for Eq.~(\ref{B0}) is given by the division of $J$ in Table~\ref{WaTable} by $S_{\mathrm{a}}$. From the pairs of $l_{\mathrm{c}}$ and $j_0$, $B_0$ is estimated using Eq.~(\ref{B0}) to be $(2.5 \pm 0.3) \times 10^{12}\,\mathrm{A\, m^{-2}\, sr^{-1}}$, with uncertainty resulting from the measurement errors of the $j_0$ and $l_{\mathrm{c}}$ values. For reference, $\bar{B}$ values reported previously for the Schottky FEG\cite{Williams2009,Humphreys1981} are in the $(2-10) \times 10^{12}\,\mathrm{A\, m^{-2}\, sr^{-1}}$ range, which is in digit agreement with the present $B_0$ value. Figure~\ref{CurrentOri} shows the curve for Eq.~(\ref{wignerorigin}) calculated using the determined $B_0$ value and the plot of the $W_{\mathrm{a}} (\bm{0}, \bm{0}) k^2 J$ for each beam listed in Table~\ref{WaTable}. The good agreement in Fig.~\ref{CurrentOri} verifies the correctness of Eqs.~(\ref{wignerorigin}) and (\ref{B0}) under a condition in which the influence of the amplitude and phase distributions in the aperture is small enough to be ignored.

The reason why not small phase shifts in the aperture as shown in Fig.~\ref{Airyresult}(d) do not affect the calculation in Eq.~(\ref{Wa00woapprox}) could be that $e^{-i k (X(\frac{\bm{\mu}}{2})- X(-\frac{\bm{\mu}}{2}))}\simeq 1$ tends to hold in many cases. It is well known in the field of electron microscopy that the main factors of the geometric aberration in the TEM electromagnetic lenses are the defocus aberration, third-order spherical aberration, and two-fold astigmatism. The phase shifts induced by the defocus and spherical aberrations are axially symmetric, and that induced by the two-fold astigmatism is two-fold rotationally symmetric with respect to the optical axis\cite{Erni2015}. Therefore, as far as the TEM lenses are aligned carefully so that the optical axis is located near the aperture center, Eqs.~(\ref{Wa00woapprox})--(\ref{wignerorigin}) are available for the $B_0$ measurements based on the Airy pattern analysis.

The significance of Eq.~(\ref{B0}) is that $B_0$ is expressed using $l_{\mathrm{c}}$ and $j_0$, which are intrinsic beam characteristics rather than the emittance, which is affected by the measurement conditions. Therefore, if $j_0$ and $l_{\mathrm{c}}$ are correctly measured by any method, for example by using an electronic biprism, then $B_0$ can be determined using Eq.~\ref{B0}. Interestingly, a similar expression for the mean brightness was proposed previously as $\bar{B}_{\mathrm{p}}=4\pi j_0 l_{\mathrm{c}}^2/\lambda^2$\cite{Pozzi1987}. This formula is only valid when an illumination aperture with a radius of $1⁄l_{\mathrm{c}}$ is used for the $j_0$ measurements, unlike Eq.~(\ref{B0}), which is free from that aperture size because of much smaller SA than beam diameters. Therefore, if we substitute the same values of $l_{\mathrm{c}}$ and $j_0$ with Eq.~(\ref{B0}), $\bar{B}_{\mathrm{p}}$ with twice the value of $B_0$ is given \cite{Hatanaka2021}, which is contrary to the general trend noted previously, where $\bar{B} < B_0$. Even if $\bar{B}$ is measured correctly by  by other methods without using the formula, in addition to the systematic errors caused by the trend, the value fluctuates depending on the TEM and the aperture used, as mentioned earlier. The derived Eq.~(\ref{B0}) is an important result that enables electron emitter performance evaluation with high precision and accuracy without being affected by differences in the optical systems.

\section{CONCLUSION} 
We have developed a reconstruction method for the density matrix and the Wigner function of electron beams based on Airy pattern intensity analysis. The reconstruction method of the density matrix can be applied to decoherence measurement of inelastically scattered electrons. It is reported that the mutual coherence of inelastically scattered electrons due to plasmon excitation decays rapidly before reaching 10~nm, and the decay curve is not represented by Gaussian distribution\cite{Schattschneider2005}. Such changes of electron states via inelastic scattering are reflected to the Wigner function. If electrons are in nonclassical states, the Wigner function can include negative values. Using this characteristic of the Wigner function, whether the scattered electrons are in nonclassical states or not will be investigated. 

In a previous study, the coherence of electrons inelastically scattered by bulk plasmons was measured using an electron biprism and an energy filter\cite{Potapov2006}. As discussed in \cite{Hatanaka2021}, coherence measurements performed using a biprism present difficulties for poorly coherent electrons such as inelastically scattered electrons. This is because the beam is partly shielded by the biprism itself, settled on the optical axis. However, the coherence measurement of inelastically scattered electrons is possible using a specially-fabricated small aperture\cite{Hatanaka2021} and the energy filter to select the energy-loss electrons.


As a result of reconstructing the electron states for beams passing through the aperture, we derived a formula to calculate the axial brightness and then determined the axial brightness of the Schottky FEG precisely and accurately without being affected by differences in the optical system. The ability to estimate axial brightness will be beneficial for precise comparison of the performances of various types of electron emitters including photocathodes\cite{Kuwahara2021,Hatanaka2023}. Monitoring the degradation process of the emitter performance under various conditions involving vacuum pressure, dark current, and so on should be greatly helpful to obtain the guideline for effective developments of high-performance emitters. The efficiency in developing advanced emitters should be maximized if the present method will be successfully applied not only in TEMs but also in simple vacuum chambers without lenses.
\\

We are grateful for Prof.~Y.~Nakata of Osaka University and Prof.~R.~Nishi of Fukui University of Technology for invaluable comments and discussion. This work was partially supported by JSPS KAKENHI grant no. 22K18974, 20K15174, and 26286049.

\bibliography{aps}

\end{document}